\documentclass[10pt,showpacs,twocolumn,nofootinbib,floatfix,aps,prd]{revtex4-1}
\usepackage{amsmath} \usepackage{graphicx} \usepackage{amsfonts}
\usepackage{array} \usepackage{amsthm} \usepackage{bm}
\usepackage{palatino} \usepackage{mathpazo} 
\usepackage{supertabular}
\usepackage{multirow}
\usepackage[breaklinks]{hyperref}
\usepackage{color}

\usepackage{epstopdf}
\usepackage{graphicx}

\begin{document}

\title{Spherical Accretion onto Konoplya Zhidenko Black Holes: Sonic Point Analysis and Fluid Dynamics}

\author{M H Waheed}\email{mhammad.bsh23ceme@student.nust.edu.pk}
\affiliation{Department of Basic Science and Humanities, College of Electrical and Mechanical Engineering, NUST, H-12, Islamabad, Pakistan}

\author{M Z A Moughal}\email{moughalzubair@gmail.com}
\affiliation{Department of Basic Science and Humanities, College of Electrical and Mechanical Engineering, NUST, H-12, Islamabad, Pakistan}

\begin{abstract}
	We investigate the spherical accretion onto a static Konoplya Zhidenko black hole and examine how the deformation of the spacetime affects the dynamics of the accreting matter. By applying the conservation laws of particle number and energy-momentum, we formulate the accretion problem and derive the corresponding Hamiltonian describing the fluid motion. The critical points and sonic transitions are then analyzed for ultra stiff, ultra relativistic, radiation, and sub relativistic fluids. We study the dependence of the critical point structure and flow trajectories on the KZ deformation and compare the results with those obtained for asymptotically safe, Schwarzschild MOG, and Reissner Nordström black holes. The comparison shows that the location and behavior of the critical points vary significantly with the underlying BH geometry and the equation of state of the fluid. In particular, the KZ deformation produces noticeable changes in the radial structure of the accretion flow, with different effects observed for the considered fluid models. These results provide a comparative picture of spherical accretion in different black hole spacetimes and help clarify how modifications of the background geometry influence matter dynamics in the strong field region.
\end{abstract}

\maketitle

\section{Introduction}

Black holes (BHs) are among the most important predictions of General Relativity and provide a framework for studying gravitational phenomena in the strong field regime. The Schwarzschild solution gives the simplest picture, a static and spherically symmetric BH, while Kerr's solution extends this to the rotating case \cite{c1,c2}. Their geometry and physical behavior, horizons, singularities, particle dynamics, and gravitational effects, have all been studied at length within relativistic gravity \cite{c3,c4,c7,c8,c9}. More recently, this theoretical picture has been brought into direct contact with observation. The imaging of M87* and Sgr A*, for instance, has opened a real window onto the geometry of compact objects and made it possible to test gravitational models close to the BHs \cite{c11,c12}. Alongside this, several studies have looked at how particles move around BHs and how gravitational lensing behaves in modified gravitational backgrounds \cite{c13,c14,c16}. What emerges from this body of work is a fairly clear point, the geometry of spacetime has a strong say in how matter and radiation behave once they get close to a BH.

Accretion is one of the key processes by which a compact object builds up mass from its surroundings, and it is especially informative in the BH context, since the motion of infalling matter is shaped so directly by the background field. In effect, the way matter flows inward carries information about the spacetime geometry itself, a point discussed in some detail in relativistic treatments of accretion \cite{c17}. The idea goes back to Bondi and Hoyle, who worked out spherical accretion by stars \cite{c18}, and to Bondi's later steady-state model in the Newtonian limit \cite{c19}. Michel then carried this into General Relativity, describing steady flow onto BHs \cite{c20}. 
Shapiro discussed the accretion with physical conditions, stellar winds, magnetic fields, magnetized interstellar matter, among others \cite{c21,c22,c23}, and the backreaction of the accreting gas on the BH itself has also been examined \cite{c24}. Accretion of perfect fluids, scalar fields, modiefied Gravity, dark matter and dark energy accretion have been studied in \cite{c25,c25a,c25b,c25c, c26a, c26}. Some of this work has moved into higher dimensions \cite{c27,c28}, including a recent study of spherical accretion in higher-dimensional Reissner Nordström BHs \cite{c29}, and accretion with backreaction has even been looked at in cylindrically symmetric configurations \cite{c30}. Altogether, these results point to a close link between the accretion flow's properties and both the BH's geometry and the nature of the accreting matter. One of the more interesting features of relativistic accretion is the appearance of a critical, or sonic, point, the location where the flow shifts from subsonic to supersonic. Studying this transonic behavior tells us a good deal about the physical conditions matter experiences as it approaches the BH. Das, for example, examined transonic spherical accretion onto Schwarzschild BHs and highlighted just how central the critical point is to the overall flow \cite{c29,c32}. Hamiltonian methods have also proven useful here, offering a fairly clean way to locate critical points and work out the resulting flow structure \cite{c34}, and this approach has since been applied to accretion around regular BHs and BHs in f(t) gravity \cite{c34,c35,c36}. More broadly, accretion in different gravitational theories has been explored through both analytical and numerical methods \cite{c37, c38, c39, c41}, work that naturally raises the question of how accretion behaves in other parametrically deformed geometries, which is where the present study comes in. Parametrized BH solutions are a convenient way to explore such deviations from the Schwarzschild geometry, without having to commit to any single alternative theory of gravity up front. Rezzolla and Zhidenko introduced one such parametrization for spherically symmetric BHs in metric theories of gravity \cite{c42}, and Konoplya, Rezzolla, and Zhidenko later extended it to the axisymmetric case \cite{c43}. These parametrizations describe deviations from the standard geometry through additional deformation parameters, and the resulting Konoplya Zhidenko framework has since been used to study BH properties and their possible observational signatures \cite{c44,c45}. Recent work has, for instance, looked at gravitational lensing and thermodynamic properties of Konoplya Zhidenko (KZ) BHs specifically \cite{c46}. Taken together, these studies suggest that even fairly modest changes to the BH geometry can leave a visible mark on the physics near the horizon.

With this motivation, we study steady, spherically symmetric accretion of a perfect fluid onto the static KZ BH, in which the departure from the Schwarzschild geometry is characterized by a deformation parameter $\eta$. Using the conservation laws for particle number and energy-momentum, we derive the equations governing the accretion flow and formulate the problem within the Hamiltonian framework to determine the critical points and sonic transitions. We pay particular attention to the effect of $\eta$ on the location of the sonic point and the overall flow properties, with the Schwarzschild solution recovered in the limit $\eta=0$. We consider several types of fluids, namely ultra-stiff, ultra-relativistic, radiation, and sub-relativistic fluids, and examine how their accretion behavior changes in the deformed spacetime. To place the KZ results in a broader context, we compare the corresponding critical points and flow properties with those obtained for asymptotically safe, Schwarzschild-MOG, and Reissner Nordström (RN) BHs. These comparisons allow us to identify the features of spherical accretion that are sensitive to the underlying spacetime geometry and to clarify how the KZ deformation influences matter dynamics in the strong-field region.

\section{Static Konoplya Zhidenko Black Holes}

In this work, we study accretion onto a static KZ BH, as proposed by Kanoplya and Zhidenko \cite{c43}. This BH is characterized by a parametric deformation of the geometry, constructed via a power-series expansion, in order to incorporate alternative theories of gravity, as given in \cite{c45}. This approach yields a highly general framework, and for this reason the KZ BH is also referred to as a deformed form of the Schwarzschild BH. In this framework, the mass function is likewise expressed as a power-series expansion
\begin{equation}
	M \rightarrow M + \frac{1}{2}\sum_{i=0}^{\infty} \frac{\eta_i}{r^i}. \label{eq1}
\end{equation}
As a result, the metric function, which in the Schwarzschild case takes the standard form, is modified to
\begin{equation}
	f(r) \rightarrow 1 - \frac{2M}{r} - \sum_{i=0}^{\infty} \frac{\eta_i}{r^{i+1}}. \label{eq2}
\end{equation}
In the given expression, if we set $\eta_0 = 0$, the metric reduces to the Schwarzschild BH. Parametrized post Newtonian (PPN) experiments, including Cassini spacecraft observations \cite{hh}, along with other solar system experiments constrain the first order deformation parameter to $\frac{\eta_1}{2} \lesssim 2.3 \times 10^{-4}$. For this reason, we consider $\eta_1 = 0$, while retaining the second order term for our observational analysis, such that \(\eta_i = \eta \, \delta_{i2}\) \cite{hh1}, With this choice, Eq. (\ref{eq1}) can be rewritten as
\begin{equation}
	M \rightarrow M + \frac{\eta}{2r^2}.
\end{equation}
Using this deformed mass function, we can now write the line element for the KZ BH as follows
\begin{equation}
	ds^2 = f(r)dt^2- f^{-1}(r) dr^2- r^2 d\Omega^2,
\end{equation}
where $f(r)=1 - \frac{2M}{r} - \frac{\eta}{r^3}$ and $d\Omega^2=d\theta^2+\sin^2\theta\,d\varphi^2$. The Static KZ framework was developed from a modified gravity theory and describes the possible deformation of the BH geometry on a large scale. This deformation is represented by the parameter \(\eta\) in the metric. Within this framework, we can study the gravitational lensing, BH shadow, and the trajectories of photons and particles near the event horizon. The radius of the SKZ BH can be written as
\begin{equation}
	r_h=\frac{1}{3}\left(C+\frac{4M^2}{C}+2M\right).
\end{equation}
where
\begin{equation*}
	C=\sqrt[3]{8M^3+\frac{27}{2}\eta+
		\frac{3\sqrt{3}}{2}\sqrt{\eta(27\eta+32M^3)}}.
\end{equation*}
In the limiting case \(\eta=0\), the horizon radius reduces to \(r_h=2M\), recovering the standard event horizon of the Schwarzschild BH. The geometric structure of the KZ BH can be characterized by curvature invariants, such as the following \cite{hh2}
\begin{align*}
	I_1 &= g^{\mu \nu} R_{\mu \nu} = -\frac{2\eta}{r^5}, \\
	I_2 &= R^{\mu \nu} R_{\mu \nu} = \frac{26\eta^2}{r^{10}}, \\
	I_3 &= R^{\mu \nu \rho \sigma} R_{\mu \nu \rho \sigma}
	= \frac{8\left(6M^2 r^4 + 20M r^2 \eta + 23 \eta^2\right)}{r^{10}} .
\end{align*}
The curvature invariants help us to understand the geometric structure and singularity properties of the KZ BH. The Ricci scalar \(I_1\) and Ricci tensor invariant \(I_2\) depend on the deformation parameter \(\eta\), which shows the effect of the KZ deformation on the spacetime curvature. The Kretschmann scalar \(I_3\) gives the curvature of the spacetime and becomes divergent as \(r\rightarrow0\), showing the presence of a curvature singularity at the origin. For \(\eta=0\), these invariants reduce to the corresponding Schwarzschild values, showing that the KZ BH reduces to the Schwarzschild spacetime when the deformation parameter vanishes.

\subsection*{Formulation of Spherical Accretion}

To study the spherical accretion of matter onto a compact object, the accretion matter is considered as a perfect fluid. The velocity, pressure, and energy density of the fluid are used to describe the accretion flow and to calculate the accretion rate around the compact object. If \(n\) denotes the number of particle in the perfect fluid and \(u^\mu\) represents its four-velocity, then the corresponding particle current density can be written as \(J^\mu=nu^\mu\). According to the law of conservation, expressed mathematically as
\begin{equation}
	\nabla_\mu J^\mu=0, \label{eq3}
\end{equation}
where \(\nabla_\mu\) denotes the covariant derivative. Similarly, according to the law of conservation of energy, mathematically expressed as
\begin{equation}
	\nabla_\mu T^{\mu\nu}=0, \label{eq4}
\end{equation}
where \(T^{\mu\nu}\) represents the energy momentum tensor of the fluid, given by
\begin{equation}
	T^{\mu\nu}=(\rho+p)u^\mu u^\nu+pg^{\mu\nu},
\end{equation}
where \(\rho\) is the energy density, \(p\) is the pressure, \(g^{\mu\nu}\) represents the metric tensor, and \(u^\mu=\) are the four vector velocity.

For Bondi-type accretion, we assume that the fluid flow is steady and spherically symmetric. Therefore, the fluid quantities do not depend explicitly on time or angular coordinates and can be taken as functions of the radial coordinate $(r)$. By using these assumptions in Eqs. (\eqref{eq3}) and (\eqref{eq4}), the equations can be reduced to the following form:
\begin{align}
	r^2nu'&=c_1, \label{eq5}\\
h^2\left[r^3(u')^2+r^3-2Mr^2-\eta\right]&=c_2r^3.\label{eq6}
\end{align}
where \( h \) denotes the specific enthalpy of the fluid, The quantities \( c_1 \) and \( c_2 \) are constants of integration that arise from the conservation equations and characterize the accretion flow.

\subsection*{Sonic Point Analysis}

The sonic point (or critical point) corresponds to the location where the inward radial velocity of the accreting fluid becomes equal to the local speed of sound. At this radius the accretion flow undergoes a transition from subsonic to supersonic motion. The behavior of the flow at this point is important because physically acceptable accretion solutions must pass smoothly through the sonic point. The square of the local speed of sound is defined as
\begin{equation}
	a^2 = \left( \frac{u'}{u_t} \right)^2,
	\label{h3}
\end{equation}
where \(a\) represents the sound speed of the fluid. If the pressure of the fluid depends only on its density, the fluid is said to obey a barotropic equation of state. In this case the specific enthalpy becomes a function of the particle number density only, i.e., \(h=h(n)\). Under this assumption the differential relation between the enthalpy and particle density takes the form
\begin{equation}
	\frac{dh}{h}=a^2\left(\frac{dn}{n}\right).
	\label{h4}
\end{equation}
Using the normalization condition of the four--velocity,
\[
u^\mu u_\mu=-1,
\]
Eq.~(\ref{eq6}) can be rewritten as
\begin{equation}
	h^2(u^t)^2=c_2.
	\label{h5}
\end{equation}
Next, we differentiate Eq.~(\ref{eq5}) with respect to the radial coordinate. This gives the relation
\begin{align}
	\frac{dn}{n}+\frac{du^r}{u^r}+\frac{2}{r}dr=0 .
	\label{h6}
\end{align}
Similarly, differentiating Eq.~(\ref{eq6}) yields
\begin{align}
	2h\,dh\,(f+(u^r)^2)+h^2\,(df+2u^r\,du^r)=0 \nonumber \\
	\frac{dh}{h}+\frac{df+2u^rdu^r}{2(f+(u^r)^2)}=0 .
	\label{h7}
\end{align}
Substituting Eq.~(\ref{h6}) into Eq.~(\ref{h4}) gives
\begin{align}
	\frac{dh}{h}=-a^2\left(\frac{du^r}{u^r}+\frac{2}{r}dr\right).
\end{align}
Replacing this expression in Eq.~(\ref{h7}) leads to
\begin{align}
	-a^2\left(\frac{du^r}{u^r}+\frac{2}{r}dr\right)
	=
	-\frac{df+2u^rdu^r}{2\left(f+(u^r)^2\right)} .
\end{align}
After rearranging and simplifying the above expression, the radial velocity gradient of the accreting fluid becomes
\begin{align}
	\frac{du^r}{dr}=
	\frac{\dfrac{df}{dr}-\dfrac{4a^2(f+(u^r)^2)}{r}}
	{2u^r-\dfrac{2a^2(f+(u^r)^2)}{u^r}} .
	\label{h8}
\end{align}
The sonic (critical) point occurs at \(r=r_s\), where the numerator and denominator of Eq.~(\ref{h8}) vanish simultaneously.

\subsection*{(A) Condition from the Denominator}
Setting the denominator of Eq.~(\ref{h8}) equal to zero gives
\begin{align}
	2(u^r_s)^2=2a_s^2\left(f_s+(u^r_s)^2\right).
	\label{h9}
\end{align}
Solving this relation for the sound speed yields
\begin{align}
	a_s^2=\frac{(u^r_s)^2}{f_s+(u^r_s)^2},
	\label{h10}
\end{align}
while solving it for the radial velocity gives
\begin{align}
	(u^r_s)^2=\frac{a_s^2 f_s}{1+a_s^2}.
	\label{h11}
\end{align}
\subsection*{(B) Condition from the Numerator}
The numerator of Eq.~(\ref{h8}) must also vanish at the sonic point. This condition leads to
\begin{align}
	\frac{df}{dr}\bigg|_{r_s}=
	\frac{4a_s^2\left(f_s+(u^r_s)^2\right)}{r_s}.
\end{align}
For the KZ BH metric function \(f(r)=1-\frac{2M}{r}-\frac{\eta}{r^3}\), its radial derivative becomes
\begin{align}
	\frac{2M}{r_s^2}+\frac{3\eta}{r_s^4}
	=
	\frac{4a_s^2\left(f_s+(u^r_s)^2\right)}{r_s}.
\end{align}
Substituting Eq.~(\ref{h11}) into the above relation gives
\begin{align}
	\frac{2M}{r_s^2}+\frac{3\eta}{r_s^4}
	=
	\frac{4a_s^2\left(f_s+\frac{a_s^2 f_s}{1+a_s^2}\right)}{r_s}.
	\label{h12}
\end{align}
Finally, solving the above equation for the sound speed at the sonic point yields
\begin{align}
	a_s^2=
	\frac{-7 \eta \pm\sqrt{C}-10 M r^2+4 r^3}{16 \left(\eta +2 M r^2-r^3\right)}
	\label{h13}
\end{align}
where
\begin{align*}
	C=-47 \eta ^2-28 M^2 r^4-4 M \left(4 r^5+29 \eta  r^2\right)+16 r^6+40 \eta  r^3
\end{align*}
The value \(a_s^2\) represents the square of the sound speed evaluated at the sonic radius. This quantity plays an important role in determining the properties of the accretion flow and is often used to construct plots describing the variation of the sound speed at the critical point.

\section{Test Fluids}

To investigate the accretion behavior of matter around the BH, we consider a simple class of fluids described by a linear equation of state. In this model the pressure is proportional to the energy density and is written as
\begin{align}
	p = k e , \label{eqh0}
\end{align}
where \(p\) represents the pressure of the fluid, \(e\) denotes the energy density, and \(k\) is a dimensionless state parameter satisfying \(0<k\leq1\). This type of relation is commonly used to model test fluids in relativistic accretion studies because it provides physically meaningful description of the fluid properties.

For the fluid flow considered here, we assume that the process is adiabatic, which implies that there is no change in entropy during the motion of the fluid. This condition can be expressed as
\begin{align*}
	T dS = 0 ,
\end{align*}
where \(S\) denotes the entropy and \(T\) is the temperature of the fluid. Under this condition, the first law of thermodynamics leads to a relation between the energy density and the particle number density. The change of the energy density with respect to the particle number density is given by
\begin{align}
	\frac{de}{dn} = \frac{p+e}{n} = h ,
	\label{h14}
\end{align}
where \(h\) is the specific enthalpy of the fluid. Integrating the above relation provides a connection between the particle number density and the energy density. This relation can be written as
\begin{align}
	n = n_s \left(\frac{e}{e_s}\right)^{\frac{1}{k+1}},
	\label{h15}
\end{align}
where \(n_s\) and \(e_s\) denote the particle number density and energy density evaluated at the sonic point. The above expression can also be written in the form
\begin{align*}
	\left(\frac{n}{n_s}\right)^{k+1} = \frac{e}{e_s}
	\quad \Rightarrow \quad
	\frac{e_s (n)^k}{n_s (n_s)^k} = \frac{e}{n}.
\end{align*}
Substituting this relation into Eq.~(\ref{h14}) yields an explicit expression for the specific enthalpy,
\begin{align}
	h = \frac{(k+1)e_s}{n_s}\left(\frac{n}{n_s}\right)^k .
	\label{h16}
\end{align}
Next, we use the integral of motion obtained earlier,
\begin{align}
	h^2\left(f+(u^r)^2\right)=c_2 ,
\end{align}
which can be written in the equivalent form
\begin{align*}
	h\sqrt{f+(u^r)^2} = \sqrt{c_2}.
\end{align*}
Substituting Eq.~(\ref{h16}) into this relation gives
\begin{align*}
	\frac{(k+1)e_s}{n_s}\left(\frac{n}{n_s}\right)^k
	\sqrt{f+(u^r)^2}
	=
	\sqrt{c_2}.
\end{align*}
After rearranging the above expression, we obtain
\begin{align*}
	n^k \sqrt{f+(u^r)^2}
	=
	\frac{\sqrt{c_2}\, n_s^{k+1}}{(k+1)e_s}.
\end{align*}
Using the conservation relation for the particle number density,
\[
n = \frac{c_1}{r^2 u^r},
\]
the above equation can be written in the form
\begin{align}
	\sqrt{f+(u^r)^2} = c_3\, r^{2k} (u^r)^k ,
	\label{h17}
\end{align}
where the constant \(c_3\) is defined as
\[
c_3 = \frac{\sqrt{c_2} n_s^{k+1}}{c_1 (k+1) e_s}.
\]
At the sonic point the flow variables satisfy additional critical conditions. In particular, the radial velocity at the sonic point is related to the metric function through
\begin{align}
	(u_s^{r})^2 = \frac{1}{4} r_s f_{s,r_s}
	= k\left(\frac{1}{4} r_s f_{s,r_s}+f_s\right).
	\label{h18}
\end{align}
This relation plays an important role in determining the properties of the flow near the critical point. To further analyze the dynamics of the accretion flow, it is convenient to introduce a Hamiltonian formulation. The Hamiltonian associated with the accretion flow can be written as
\begin{align}
	\mathcal{H} =
	\frac{f^{\,1-k}}
	{v^{2k}\, r^{4k} (1-v^2)^{1-k}},
	\label{h19}
\end{align}
where \(v\) represents the three--velocity of the radial flow in the equatorial plane. This velocity is defined as
\begin{align}
	v \equiv \frac{dr}{f\,dt}.
\end{align}
Using the definition of the four--velocity components,
\[
u^{r} = \frac{dr}{d\tau},
\qquad
u^{t} = \frac{dt}{d\tau},
\]
the square of the three--velocity can be expressed as
\begin{align}
	v^2
	=
	\left(\frac{u^r}{f\,u^t}\right)^2
	=
	\frac{(u^r)^2}{(u^t)^2}
	=
	\frac{(u^r)^2}{f+(u^r)^2}.
	\label{h20}
\end{align}
This relation connects the three--velocity of the fluid with the components of the four--velocity and the metric function, and it is particularly useful for studying the phase space structure and the dynamical behavior of the accretion flow.

\subsection*{A. Ultra-Stiff Fluid ($k=1$)}

\begin{figure}
	\centering
	\includegraphics[width=0.85\linewidth]{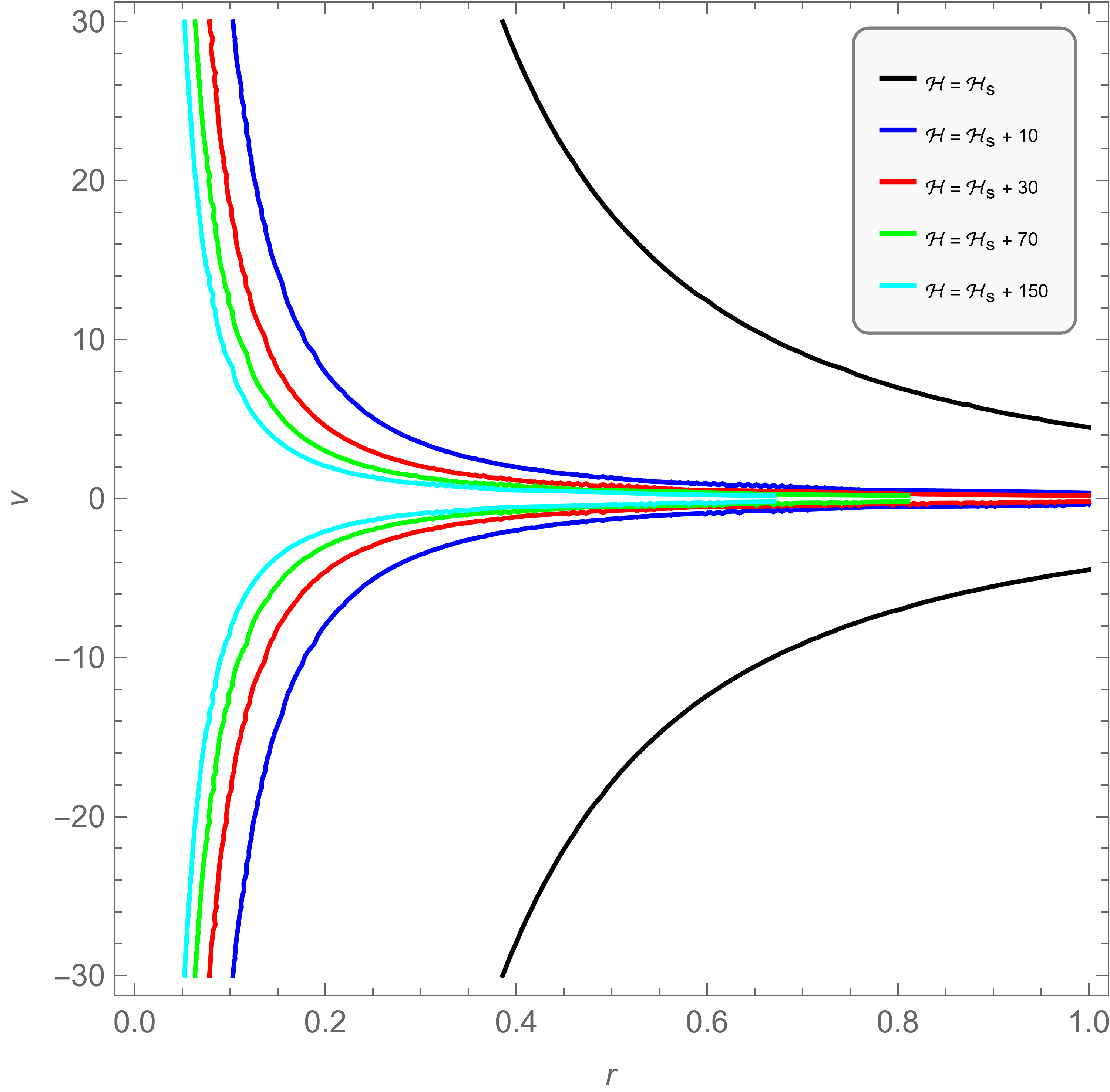}
	\caption{Plot of the radial coordinate \(r\) versus velocity \(v\) for the ultra--stiff fluid case.}
	\label{figure1}
\end{figure}

We first consider the case of an ultra--stiff fluid, which corresponds to the equation of state parameter \(k=1\). In this situation the pressure becomes equal to the energy density, \(p=e\). Such a fluid represents an extreme physical case in which the speed of sound approaches the speed of light, and it is often used in theoretical studies to examine limiting behaviors of relativistic accretion flows. Substituting \(k=1\) into Eqs.~(\ref{h17}), (\ref{h18}) and (\ref{h19}), the relation describing the radial flow of the fluid reduces to
\begin{align}
	\sqrt{f+(u^{r})^2} = c_3\, r^2 u^{r},
	\label{h21}
\end{align}
where the constant \(c_3\) takes the form
\begin{equation*}
	c_3 = \frac{\sqrt{c_2}\, n_s^{\,2}}{c_1(2)\, e_s}.
\end{equation*}
For this case the critical point coincides with the horizon of the BH, which implies that the metric function satisfies
\begin{equation*}
	f_s = 0 .
\end{equation*}
Solving the horizon condition for the Konoplya--Zhidenko spacetime gives the location of the sonic point. The positive real root of the corresponding equation can be written as
\begin{equation*}
	r_s = \frac{1}{3}\left(\sqrt[3]{C} + \frac{4M^2}{\sqrt[3]{C}} + 2M \right),
\end{equation*}
where
\begin{equation*}
	C = 8M^3 + \frac{27}{2}\eta + \frac{3\sqrt{3}}{2}\sqrt{\eta\left(27\eta + 32M^3\right)} .
\end{equation*}
The Hamiltonian describing the accretion flow can also be simplified by inserting \(k=1\) into Eq.~(\ref{h19}). In this case the Hamiltonian becomes
\begin{align}
	\mathcal{H} = \frac{f^{0}}{v^{2} r^{4}}
	= \frac{1}{v^{2} r^{4}} .
	\label{h22}
\end{align}
For the chosen parameters, the sonic point coincides with the event horizon and is located at \(r_s = r_h = 2.11208\), while the corresponding velocity and Hamiltonian values at the sonic point are \(v_s = 1, \mathcal{H}_s = 0.0502521\). The phase--space behavior of the flow for the ultra--stiff fluid case is illustrated in Fig.~\ref{figure1}, where the variation of the velocity \(v\) with respect to the radial coordinate \(r\) is shown. It can be observed that when \( |v| < 1 \), the flow becomes subsonic, which means that the flow velocity becomes smaller than the sound speed. As the value of \( r \) increases toward the outer region, the flow gradually changes from the supersonic region to the subsonic region. The transition point between these two regions is called the critical point, which occurs at \( r = r_s \).

\subsection*{B. Ultra–Relativistic Fluid ($k=\tfrac{1}{2}$)}

For the ultra–relativistic fluid, the equation of state takes the form \(p=\frac{e}{2}\),
which corresponds to the parameter value $k=\tfrac{1}{2}$. Under this assumption, the critical point condition obtained from Eq.~(\ref{h18})
reduces to
\begin{align}
	r_s f_{,r}(r_s)-4f(r_s)=0 .
\end{align}
Solving this relation provides the possible locations of the critical
radius, given by
\begin{align}
	r_s=\pm \frac{1}{2}\sqrt{\,M \pm \sqrt{12\eta + M^2}\,}.
\end{align}
Among these solutions, the physically meaningful root corresponds to the
positive value of $r_s$.

The Hamiltonian describing the dynamical system of the accretion flow is
obtained from Eq.~(\ref{h19}). For $k=\tfrac{1}{2}$, it simplifies to
\begin{align}
	\mathcal{H}=\frac{\sqrt{f}}{r^{2} v\sqrt{1-v^{2}}}.
\end{align}
This quantity remains constant along the trajectories in the phase space
$(r,v)$ and characterizes the behavior of the fluid motion around the
BH.

\begin{figure}
	\centering
	\includegraphics[width=0.85\linewidth]{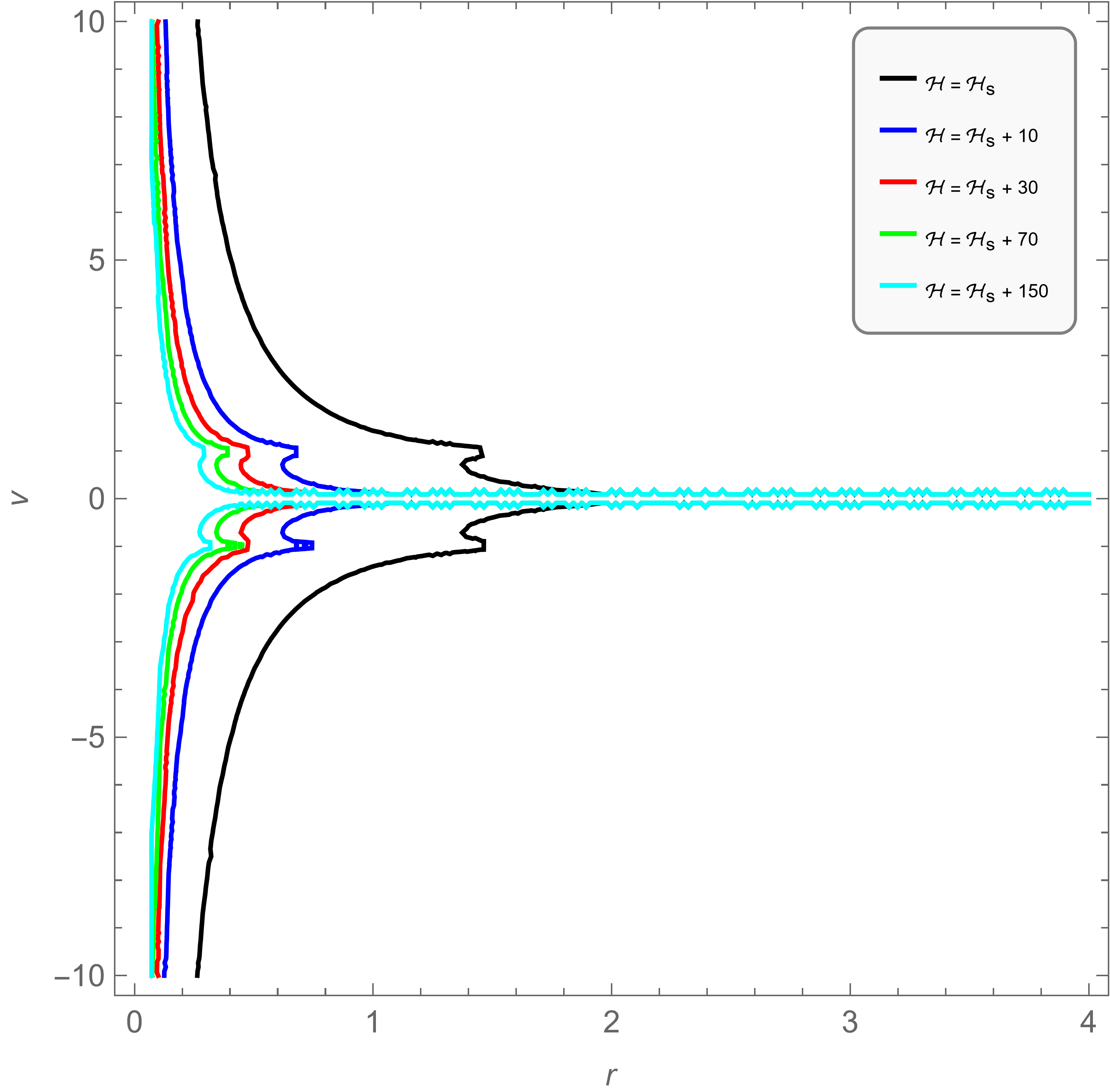}
	\caption{Phase-space plot of $r$ versus $v$ for the ultra-relativistic fluid with velocity range $-10 \leq v \leq 10$.}
	\label{figure2A}
\end{figure}

\begin{figure}
	\centering
	\includegraphics[width=0.85\linewidth]{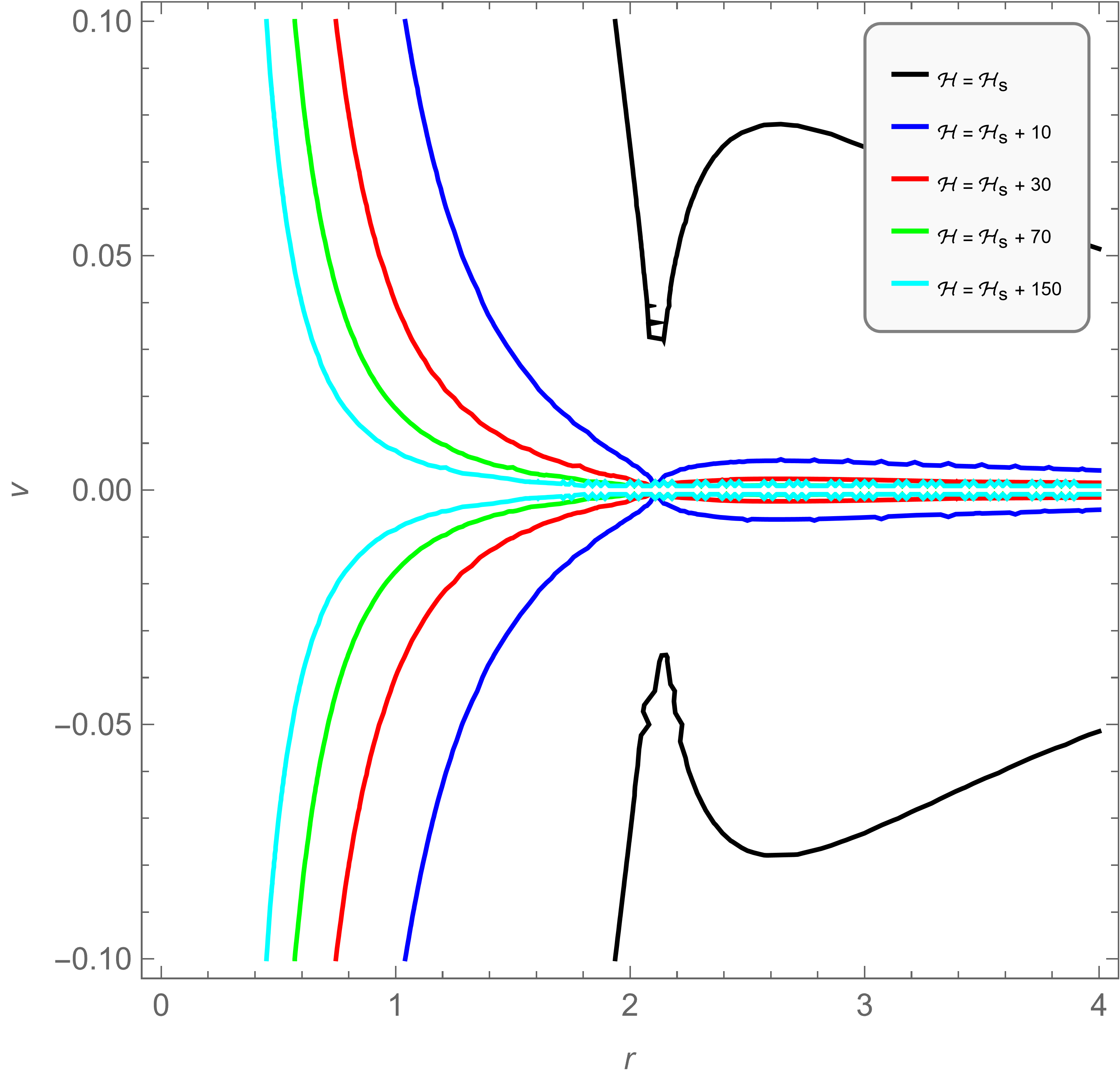}
	\caption{Phase-space plot of $r$ versus $v$ for the ultra-relativistic fluid with velocity range $-0.1 \leq v \leq 0.1$.}
	\label{figure2B}
\end{figure}

For the chosen parameter values, the critical point occurs at
$r_s=0.95469$, with the corresponding velocity $v_s=-1.49358$ and
Hamiltonian value $\mathcal{H}_s=0.0502521$. The phase–space structure
of the flow is illustrated in Figs.~\ref{figure2A} and \ref{figure2B}.
Figure~\ref{figure2A} shows the global behavior for a wider velocity
range $-10 \leq v \leq 10$, whereas Fig.~\ref{figure2B} focuses on a
narrow region near the origin to highlight the local behavior of the
trajectories. The positive and negative velocity regions in the graph represent two possible fluid motions around the BH. The positive velocity corresponds to the outflow of the fluid, whereas the negative velocity represents the inflow motion, which leads to the accretion process near the BH. The phase portrait confirms the existence of transonic accretion behavior, where the fluid evolves smoothly from a subsonic state at large radial distance to a supersonic state near the BH horizon through the critical point.

\subsection*{C. Radiation Fluid ($k=1/3$)}

\begin{figure}[h]
	\centering
	\includegraphics[width=0.85\linewidth]{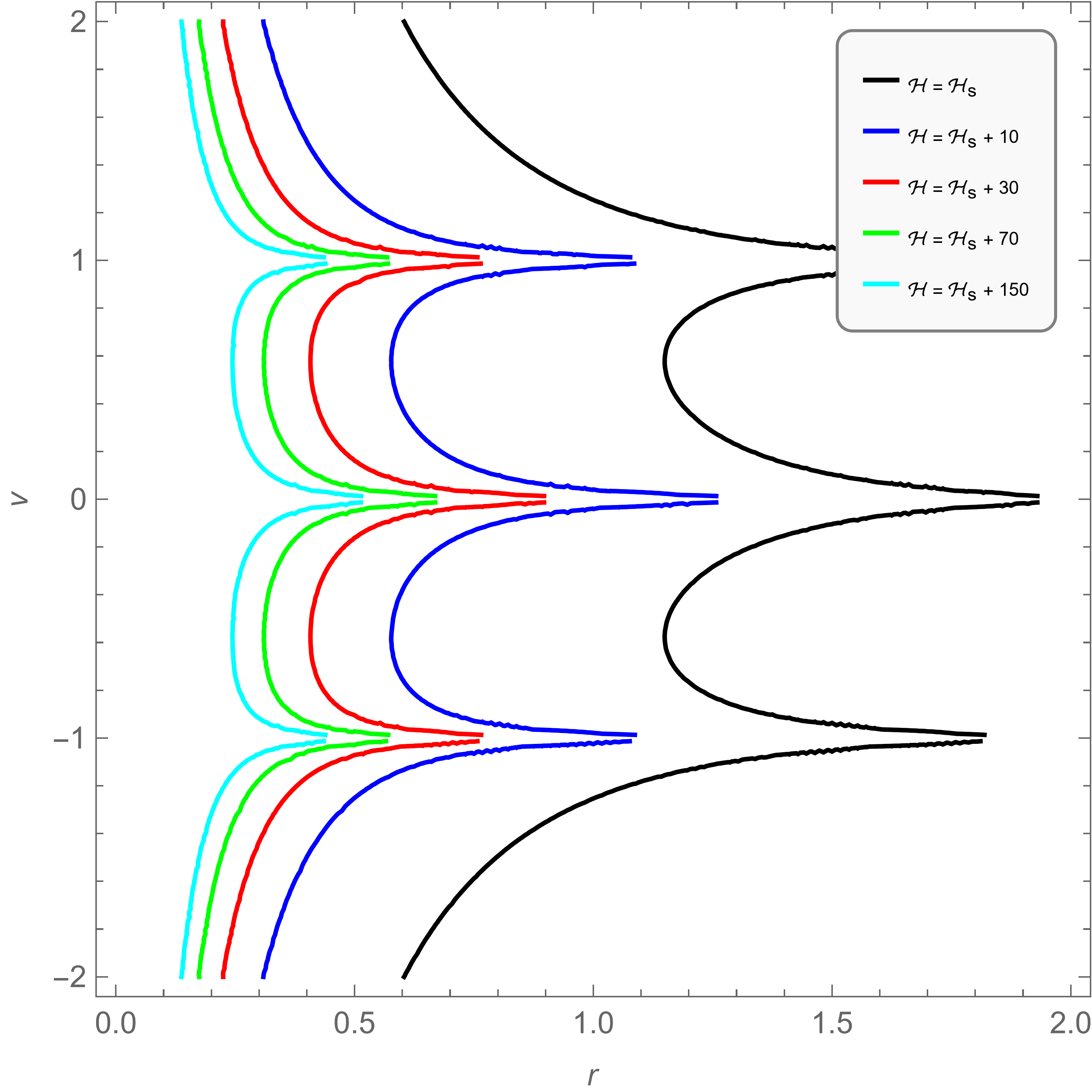}
	\caption{Plot of $v$ versus $r$ for the radiation fluid.}
	\label{figure3}
\end{figure}

For the radiation fluid case, the equation of state is taken as \(p=\frac{e}{3}\), which corresponds to the relativistic radiation-dominated matter. Substituting the value $k=\frac{1}{3}$ into the general relations derived in the previous section allows us to obtain the conditions that govern the accretion flow for this particular fluid. Using the sonic point condition obtained from the isothermal relation, the critical point equation reduces to
\[
r_s f_{s,r_s}-2f_s=0,
\]
which determines the location of the sonic radius $r_s$. This radius represents the point where the radial flow velocity of the fluid becomes equal to the local sound speed. Similarly, by inserting $k=\frac{1}{3}$ into the general dynamical relation for the radial motion, the equation describing the behavior of the radial velocity becomes
\[
\left(f+(u^r)^2\right)^3=c_3^6 r^4 (u^r)^3.
\]
This relation connects the radial velocity component $u^r$ with the metric function $f(r)$ and characterizes the evolution of the accreting flow. The Hamiltonian describing the dynamical system for the radiation fluid takes the form
\[
\mathcal{H}=\frac{f^{2/3}}{r^{4/3} v^{2/3}(1-v^2)^{2/3}},
\]
which is useful for analyzing the phase-space structure of the accretion flow. 
Here $v$ represents the three-velocity of the fluid measured in the equatorial plane. Solving the critical point equation gives the sonic radius as
\[
r_s= \frac{1}{3}\left(
M-\frac{C^{1/3}}{2^{2/3}}-\frac{2^{2/3}M^2}{C^{1/3}}
\right),
\]
where
\[
C=-81\eta+9\sqrt{81\eta^2+8\eta M^3}-4M^3 .
\]
For the chosen set of parameters, the numerical values of the physical quantities at the sonic point are obtained as \( r_s=3.12777, u_s=0.0765349, v_s=0.0167319, \mathcal{H}_s=1.64172\). The behavior of the accretion flow in the phase space is illustrated in Fig.~\ref{figure3}, where the relation between the radial coordinate $r$ and the velocity $v$ is plotted for the radiation fluid case. The given plot illustrates the dynamical behavior of fluid flow around the BH geometry. For large values of the radial coordinate \(r\), the fluid velocity approaches small values, indicating slow motion far away from the BH. As the fluid moves closer to the BH, the velocity increases due to the strong gravitational attraction, and the flow eventually enters the supersonic regime. The behavior of the flow changes with increasing values of the Hamiltonian. The phase portrait confirms the existence of stable transonic fluid flow around the BH geometry, where the fluid smoothly evolves from the subsonic region to the supersonic region through the critical point.

\subsection*{D. Sub-Relativistic Fluid ($k=1/4$)}

\begin{figure}[h]
	\centering
	\includegraphics[width=0.85\linewidth]{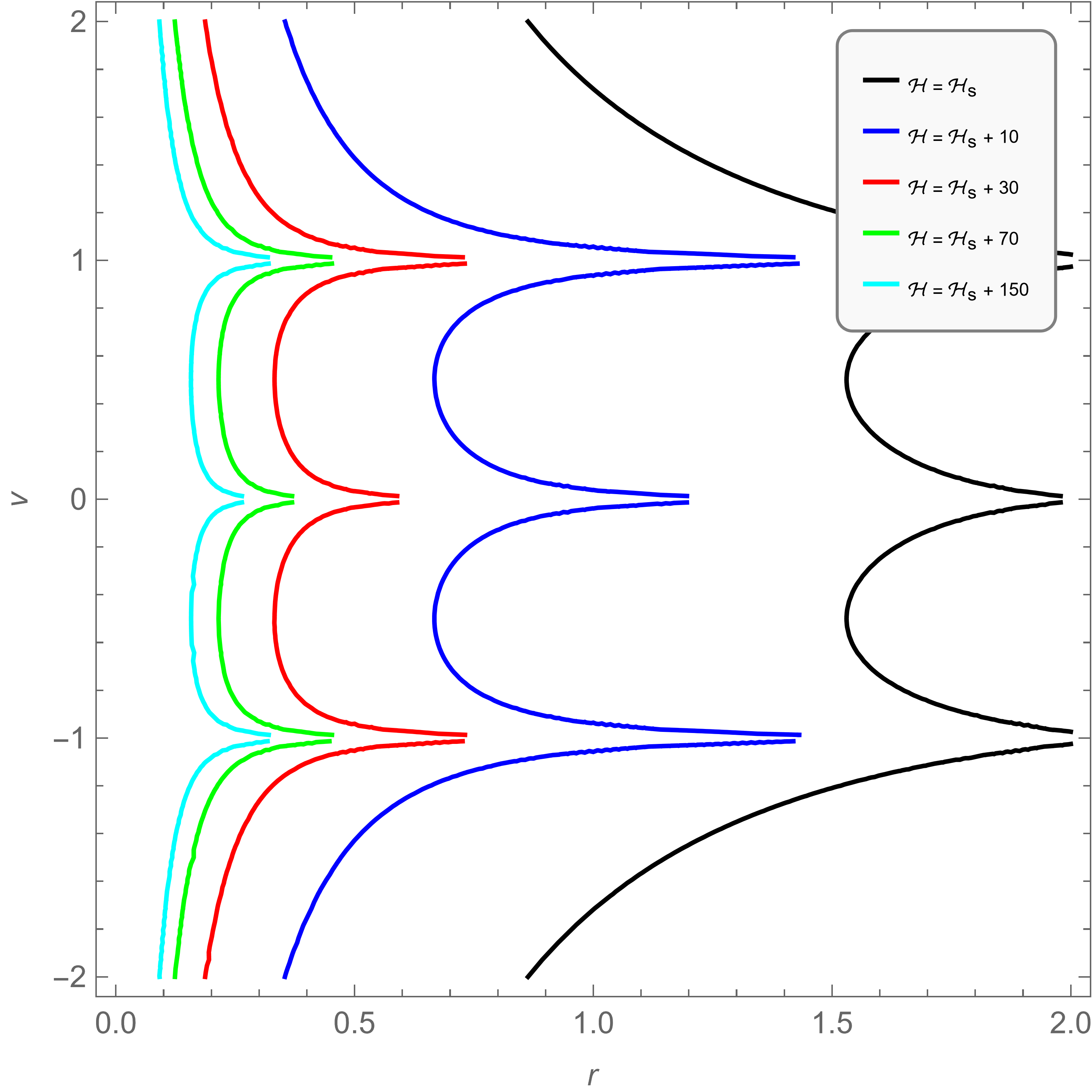}
	\caption{Plot of $v$ versus $r$ for the sub-relativistic fluid ($k=1/4$).}
	\label{fig:subrelativistic}
\end{figure}

For the sub-relativistic case, the equation of state parameter is taken as $k=1/4$. 
Substituting this value into the critical point condition leads to the following relation
\begin{align}
	3r_s f_{s,r_s}-4f_s=0 .
\end{align}
Solving the above equation gives the corresponding sonic radius
\begin{align}
	r_s= \frac{1}{6} \left(
	7M - C^{1/3} - \frac{49 M^2}{C^{1/3}}
	\right),
\end{align}
where
\begin{align}
	C= -351\eta + 3\sqrt{39}\,\sqrt{351\eta^2 + 686\eta M^3} - 343M^3 .
\end{align}
Using the general relation obtained earlier for the fluid flow, the corresponding expression for the radial velocity becomes
\begin{align}
	\left(f+(u^r)^2\right)^2=c_3^4 r^2 (u^r).
\end{align}
The Hamiltonian describing the dynamical behavior of the accretion flow for this fluid configuration takes the form
\begin{align}
	\mathcal{H}=\frac{f^{3/4}}{r\sqrt{v}(1-v^2)^{3/4}} .
\end{align}
For the chosen parameter values, the numerical results at the sonic point are obtained as \(r_s=3.62375, u_s^r=2.23029, v_s=0.919143, \mathcal{H}_s=0.62636\). The corresponding phase-space behavior of the flow is illustrated in Fig.~\ref{fig:subrelativistic}. The plot shows the variation of the radial velocity $v$ with respect to the radial coordinate $r$ for the sub-relativistic fluid. The behavior of a sub-relativistic fluid is similar to that of radiation fluid flow; however, it differs in that the velocity increases at smaller radii more slowly compared to radiation fluid.

\subsection*{E. Black Hole's Accretion Rate}
\begin{figure}[h]
	\centering
	\includegraphics[width=0.85\linewidth]{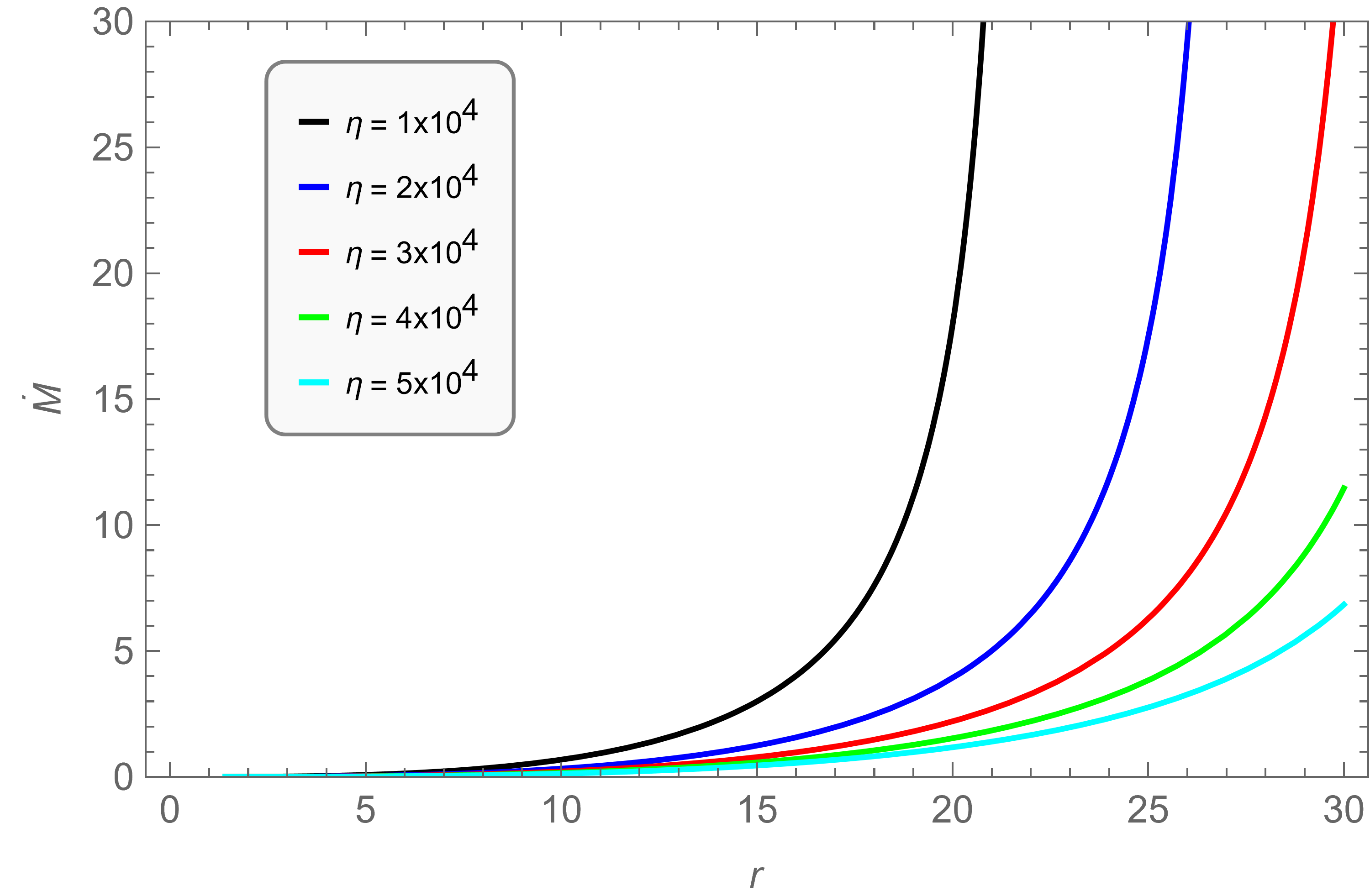}
	\caption{Variation of the BH accretion rate as a function of the radial coordinate $r$.}
	\label{figure5}
\end{figure}

The accretion rate describes the rate at which matter is absorbed by the BH due to the inward flow of the surrounding fluid. For a steady and spherically symmetric accretion flow, the mass accretion rate can be expressed as
\begin{align}
	\dot{M}=4\pi A_1 (e+p).\label{eqh1}
\end{align}
For an isothermal fluid, using the equation of state which given by Eq. (\ref{eqh0}), the Eq. (\ref{eqh1})  becomes
\[
\dot{M}=4\pi A_1 e (1+k).
\]
The energy density of the fluid can be written in terms of the integration constants and the metric function as
\begin{align}
	e=\frac{A_0^2 r^4-4A_1^4}{4A_1^2 r^4 f(r)} .
\end{align}
Substituting this expression into the accretion rate formula leads to
\begin{align}
	              \dot{M}=-\frac{\pi (k+1)\left(4A_1^4-A_0^2 r^4\right)}
	{A_1 r\left(-\eta-2Mr^2+r^3\right)} .
\end{align}
From this expression, it is evident that the accretion rate becomes singular when the denominator vanishes. The corresponding radial position of this singularity is obtained from
\begin{align}
	r=&\frac{1}{3}\left(\sqrt[3]{\frac{27\eta}{2}+\frac{3}{2}\sqrt{3}\sqrt{\eta(27\eta+32)}+8}\right) \nonumber\\
	&+\left(\frac{4}{\sqrt[3]{\frac{27\eta}{2}+\frac{3}{2}\sqrt{3}\sqrt{\eta(27\eta+32)}+8}}+2\right).
\end{align}
For the numerical analysis, the parameters are chosen as $M=A_0=A_1=k=1$. 
The behavior of the accretion rate as a function of the radial coordinate is illustrated in Fig.~\ref{figure5}. The plot shows how the accretion rate changes with distance from the BH and highlights the region where the flow becomes singular.

\section{Discussion}
The study of accretion processes and sonic points plays a fundamental role in theoretical astrophysics, as these quantities provide important information about the behavior of matter in the vicinity of BHs. In this work, we have investigated the spherical accretion of isothermal test fluids onto the KZ BH and examined how the flow properties are influenced by the fluid equation of state. Furthermore, the obtained results were compared with asymptotically safe, Schwarzschild–MOG, and RN BH geometries in order to highlight the effects of the KZ spacetime on the accretion dynamics.

To explore the dependence of the flow on the equation of state parameter (k), four physically distinct fluid configurations were considered, ultra-stiff fluid, ultra-relativistic fluid, radiation fluid, and sub-relativistic fluid. For the ultra-stiff fluid, the sonic point coincides with the event horizon \((r_s=2.11208)\), where the flow attains the speed of light \((v_s=1)\) and is characterized by a Hamiltonian value of \(\mathcal{H}_s=0.0502521\). This indicates that the transition from subsonic to supersonic flow occurs precisely at the horizon. In the ultra-relativistic case, the critical point is located at \(r_s=0.95469\) with \(v_s=-1.49358\) and \(\mathcal{H}_s=0.0502521\), suggesting a markedly different transonic behavior compared to the ultra-stiff fluid despite sharing the same Hamiltonian value. For the radiation fluid, the sonic point shifts outward to \(r_s=3.12777\), where the corresponding velocities are \(u_s=0.0765349\) and \(v_s=0.0167319\), while the Hamiltonian reaches its largest value, \(\mathcal{H}_s=1.64172\). The relatively small velocity components at the critical point indicate a gradual transition through the sonic region. In contrast, the sub-relativistic fluid possesses the largest critical radius, \(r_s=3.62375\), together with \(u_s^r=2.23029\), \(v_s=0.919143\), and \(\mathcal{H}_s=0.62636\). These values show that the flow undergoes substantial acceleration before reaching the critical point.

\begin{figure}
	\centering
	\includegraphics[width=0.95\linewidth]{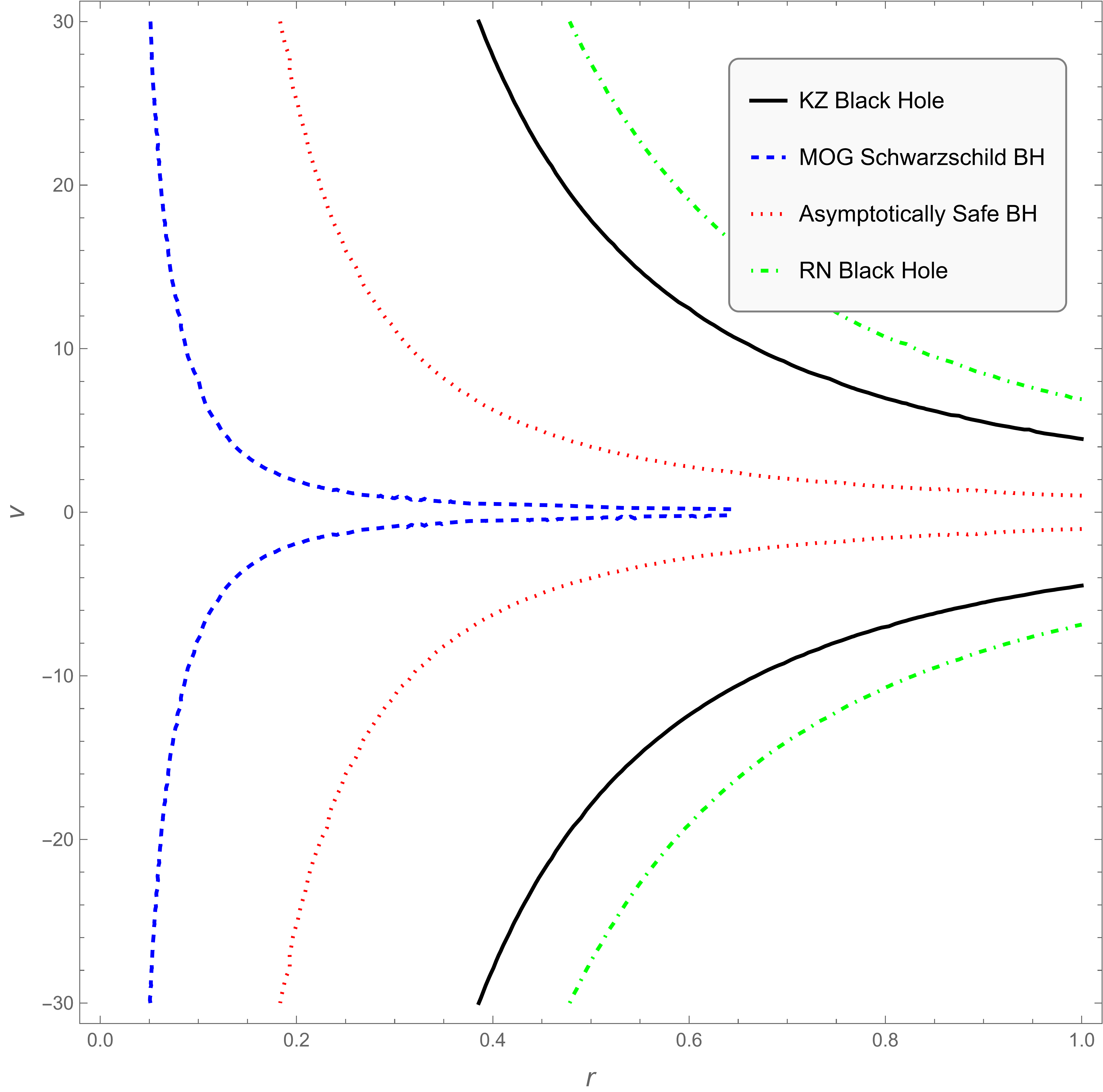}
	\caption{Comparison plot of the radial coordinate \(r\) versus velocity \(v\) for the ultra--stiff fluid case.}
	\label{comfig1}
\end{figure}

In figure (\ref{comfig1}), the comparison plot of the MOG Schwarzschild BH, asymptotically safe BH, RN BH, and KZ BH shows that the Schwarzschild MOG and asymptotically safe BHs exhibit a slower increase in the speed of sound with increasing radius compared to the KZ BH, whereas the RN BH shows the slowest variation among all the considered BHs. 

\begin{figure}
	\centering
	\includegraphics[width=0.95\linewidth]{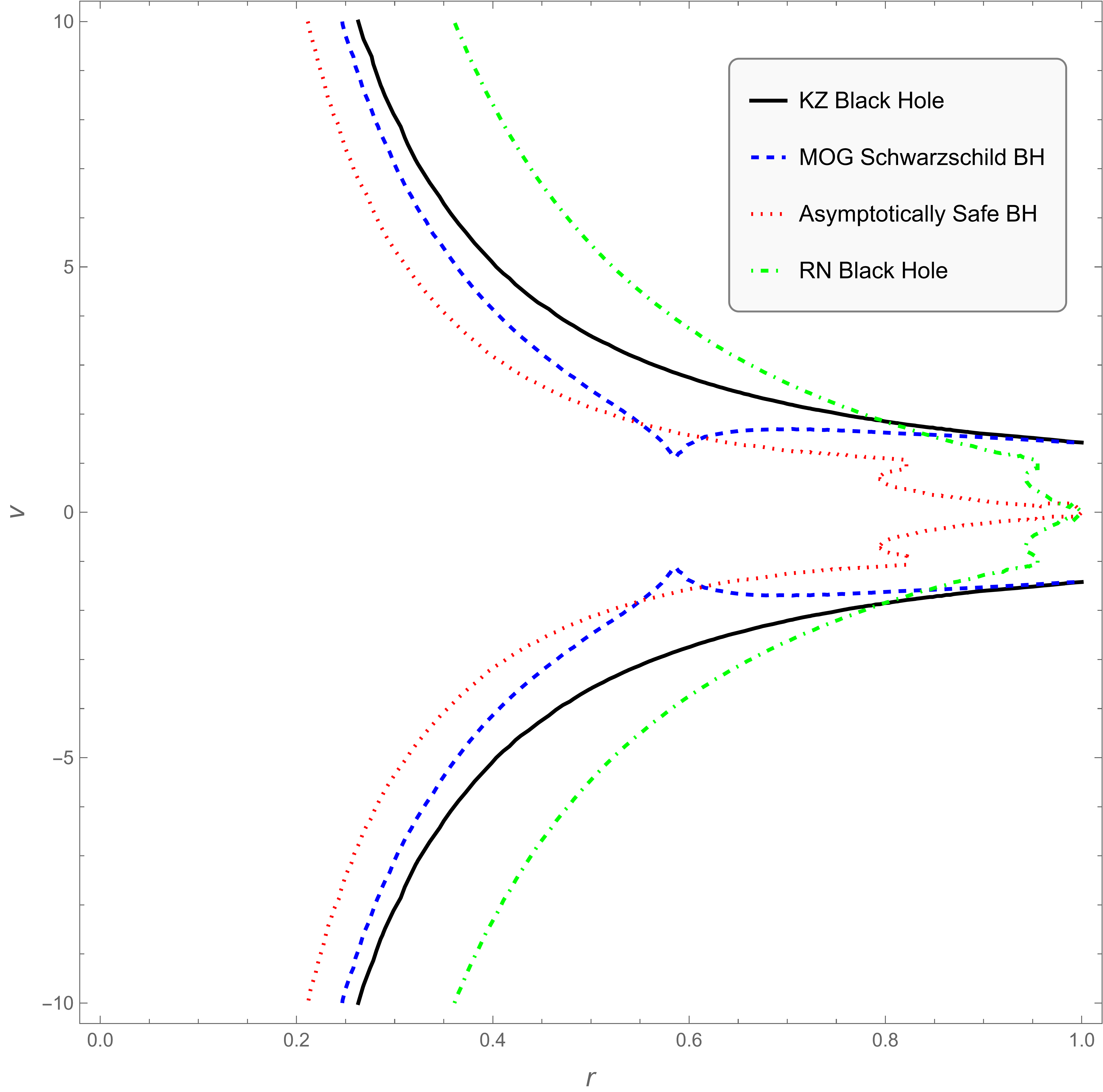}
	\caption{Comparison plot of the radial coordinate \(r\) versus velocity \(v\) for the ultra-relativistic fluid case.}
	\label{comfig2}
\end{figure}

\begin{figure}
	\centering
	\includegraphics[width=0.95\linewidth]{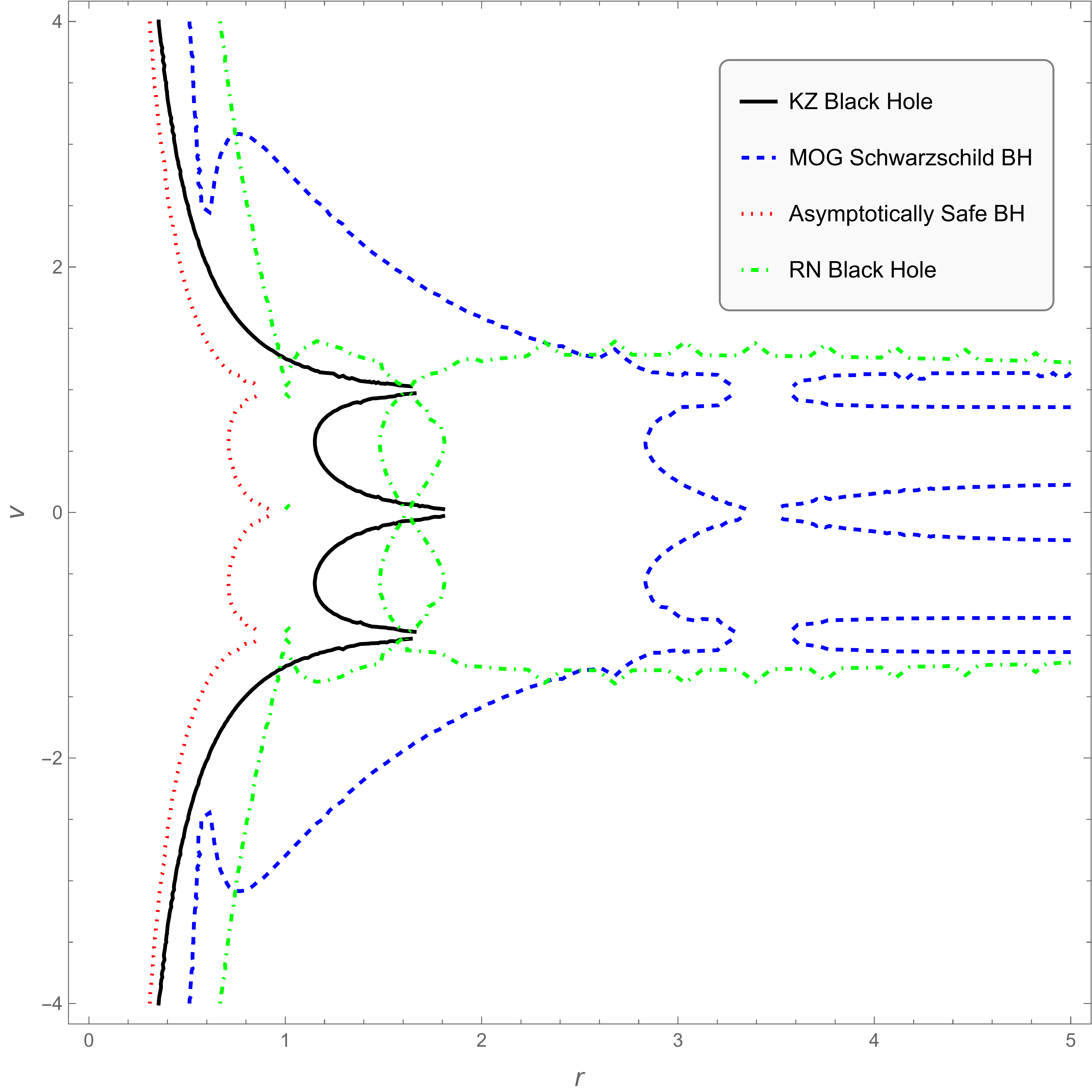}
	\caption{Comparison plot of the radial coordinate \(r\) versus velocity \(v\) for the radiation fluid case.}
	\label{comfig3}
\end{figure}

\begin{figure}
	\centering
	\includegraphics[width=0.95\linewidth]{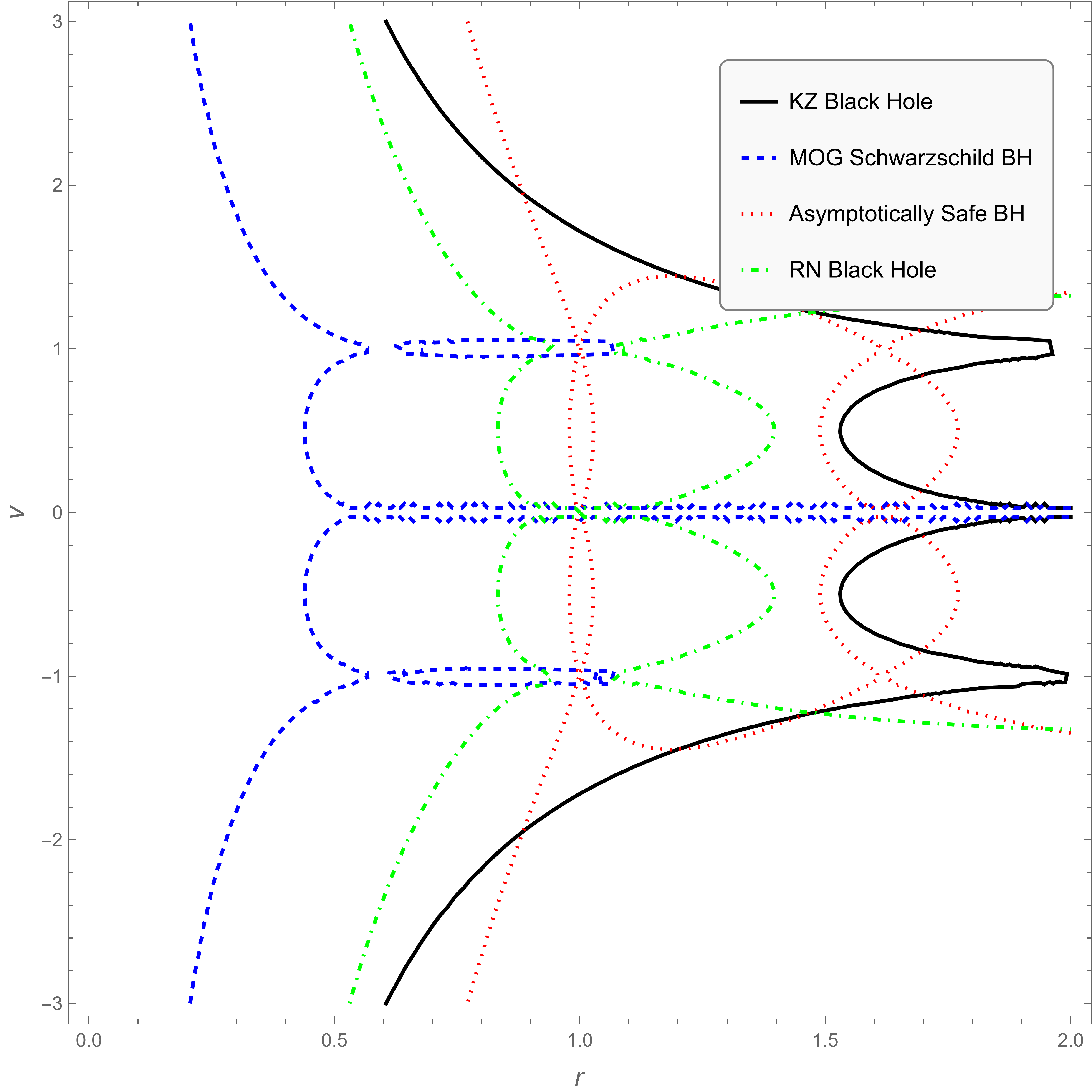}
	\caption{Comparison plot of the radial coordinate \(r\) versus velocity \(v\) for the sub-relativistic fluid case.}
	\label{comfig4}
\end{figure}

In Figure~(\ref{comfig2}), for the ultra-relativistic fluid, the KZ solution gives a relatively smooth phase-space profile, while the MOG solution remains close to it mainly in the outer region. The asymptotically safe and RN cases show larger deviations in the radial behavior. In Figure~(\ref{comfig3}), which corresponds to the radiation fluid, these differences become more pronounced. The KZ trajectories are confined to a relatively smaller radial range, whereas the MOG and RN solutions extend farther outward and exhibit additional turning structures. In contrast, the asymptotically safe BH shows a comparatively shorter radial range and smaller values of the velocity. In Figure~(\ref{comfig4}), for the sub-relativistic fluid, the KZ solution shows a distinct intermediate-radius structure, whose position and shape differ from those of the other three BH models. These results suggest that the deformation of the KZ geometry has a direct effect on the accretion flow, particularly on the radial location and structure of the critical region. The comparison also shows that this effect depends on the properties of the accreting fluid, since the differences between the BH solutions are not the same for all equations of state. For a quantitative comparison of the accretion behavior in the four BH geometries, the corresponding critical-point parameters for the different fluid models are summarized in Table~\ref{table1}.
\begin{table}[htbp]
	\centering
	\caption{Comparison of the critical radius $r_s$, critical velocity $v_s$, and Hamiltonian $H_s$ for different BH solutions under different fluid regimes.}
	\label{table1}
	\begin{tabular}{llccc}
		\hline
		\textbf{Fluid} & \textbf{Black Hole} & $r_s$ & $v_s$ & $\mathcal{H}_s$ \\
		\hline
		 
		 \multirow{4}{*}{Ultra-stiff}
		 & KZ & 2.11208 & 1.00000 & 0.0502521 \\
		 & Schwarzschild--MOG & 0.27523 & 1.00000 & 174.204 \\
		 & Asymptotically Safe & 1.00000 & 1.00000 & 1.00000 \\
		 & RN & 2.61803 & 1.00000 & 0.02129 \\
		 \hline

		\multirow{4}{*}{Ultra-relativistic}
		& KZ & 0.95469 & -1.49358 & 0.505231 \\
		& Schwarzschild--MOG & 0.27523 & 0.70711 & 0.70633 \\
		& Asymptotically Safe & 1.00000 & 1.00000 & 2.23868 \\
		& RN & 0.81896 & 0.70714 & 1.84972 \\
		\hline
		
		\multirow{4}{*}{Radiation}
		& KZ & 3.12777 & 0.167319 & 1.64172 \\
		& Schwarzschild--MOG & 4.13746 & 0.70711 & 0.14050 \\
		& Asymptotically Safe & 1.00000 & 1.00000 & 2.92443 \\
		& RN & 2.61803 & 0.57735 & 0.137061 \\
		\hline
		
		\multirow{4}{*}{Sub-relativistic}
		& KZ & 3.63275 & 0.914943 & 0.62636 \\
		& Schwarzschild--MOG & 4.00000 & 0.70711 & 0.70633 \\
		& Asymptotically Safe & 1.00000 & 1.00000 & 0.10574 \\
		& RN & 2.61803 & 0.50000 & 0.19024 \\
		\hline

	\end{tabular}
\end{table}

\section{Declaration of Competing Interest}
The authors declare that they have no competing financial or personal interests that could have influenced the work presented in this paper. This research was carried out purely for academic purposes, with no external funding and no conflicts of interest bearing on the results or conclusions.

\section{Data availability}
No data was generated or analyzed in this study, as it is purely theoretical in nature.

\end{document}